\documentclass[10pt,conference]{IEEEtran}
\IEEEoverridecommandlockouts
\usepackage{cite}
\usepackage{amsmath,amssymb,amsfonts}
\usepackage{algorithmic}
\usepackage{graphicx}
\usepackage{subfigure}
\usepackage{textcomp}
\usepackage{xcolor}
\usepackage{array}
\usepackage{xspace}
\usepackage{multirow}
\usepackage[hidelinks]{hyperref}
\usepackage{balance}
\usepackage{lipsum}
\usepackage{tikz}

\def\BibTeX{{\rm B\kern-.05em{\sc i\kern-.025em b}\kern-.08em
    T\kern-.1667em\lower.7ex\hbox{E}\kern-.125emX}}

\newcommand\copyrightnotice[1]{
    \begin{tikzpicture}[remember picture,overlay]
    \node[anchor=south,yshift=40pt] at (current page.south) {\parbox{\dimexpr 1\textwidth-\fboxsep-\fboxrule\relax}{\footnotesize #1}};
    \end{tikzpicture}
}

\begin{document}
\setlength{\textfloatsep}{7pt}
\title{\vspace{0.5cm}CURE: \underline{C}ode-Aware Ne\underline{u}ral Machine Translation for Automatic Program \underline{Re}pair}


\author{\IEEEauthorblockN{Nan Jiang}
\IEEEauthorblockA{\textit{Purdue University}\\
West Lafayette, USA \\
jiang719@purdue.edu}
\and
\IEEEauthorblockN{Thibaud Lutellier}
\IEEEauthorblockA{
\textit{University of Waterloo}\\
Waterloo, Canada \\
tlutelli@uwaterloo.ca}
\and
\IEEEauthorblockN{Lin Tan}
\IEEEauthorblockA{
\textit{Purdue University}\\
West Lafayette, USA \\
lintan@purdue.edu}
}

\maketitle

\begin{tikzpicture}[remember picture,overlay]
    \node[anchor=south,yshift=765pt] at (current page.south) {\parbox{\dimexpr 1\textwidth-\fboxsep-\fboxrule\relax}{\centering 2021 IEEE/ACM 43rd International Conference on Software Engineering (ICSE)}};
\end{tikzpicture}

\copyrightnotice{1558-1225/21/\$31.00~\copyright~2021~IEEE. \\ DOI 10.1109/ICSE43902.2021.00107}

\newcommand{\Comment}[1]{}

\newcommand{\code}[1]{\begin{small}``\texttt{#1}"\end{small}}


\newcommand{\todoc}[2]{{\textcolor{#1}{\textbf{#2}}}}

\newcommand{\todored}[1]{{\todoc{red}{\textbf{[[#1]]}}}}
\newcommand{\todoblue}[1]{\todoc{blue}{\textbf{[[#1]]}}}
\newcommand{\todoorange}[1]{\todoc{orange}{\textbf{[[#1]]}}}
\newcommand{\todobrown}[1]{\todoc{brown}{\textbf{[[#1]]}}}
\newcommand{\todogray}[1]{\todoc{gray}{\textbf{[[#1]]}}}
\newcommand{\todopink}[1]{\todoc{purple}{\textbf{[[#1]]}}}
\definecolor{ao}{rgb}{0.0, 0.5, 0.0}
\newcommand{\todogreen}[1]{\todoc{ao}{\textbf{[[#1]]}}}
\newcommand{\todo}[1]{\todored{TODO: #1}}

\definecolor{light-gray}{gray}{0.85}
\newcommand{\hilight}[1]{\colorbox{light-gray}{#1}}

\renewcommand{\todoc}[2]{\relax}

\newcommand{\thibaud}[1]{\todogreen{Thibaud: #1}}
\newcommand{\lin}[1]{\todoblue{Lin: #1}} 
\newcommand{\nan}[1]{\todoorange{Nan: #1}}


\newcommand{\tool}{CURE\xspace} 
\newcommand{\coconut}{CoCoNuT\xspace}
\newcommand{\cocoarchitecture}{CoNuT\xspace}
\newcommand{\gptparams}{$\Theta$}
\newcommand{\nmtparams}{$\Phi$}
\newcommand{\devlike}{developer-like\xspace}


\begin{abstract}
Automatic program repair (APR) is crucial to improve software reliability. Recently, neural machine translation (NMT) techniques have been used to fix software bugs automatically. While promising, these approaches have two major limitations. Their search space often does not contain the correct fix, and their search strategy ignores software knowledge such as strict code syntax. Due to these limitations, existing NMT-based techniques underperform the best template-based  approaches.

We propose \tool, a new NMT-based APR technique with three major novelties. First, \tool pre-trains a programming language (PL) model on a large software codebase to learn \devlike source code before the APR task. Second, \tool designs a new code-aware search strategy that finds more correct fixes by focusing on 
compilable patches and patches that are close in length to the buggy code. Finally, \tool uses a subword tokenization technique to generate a smaller search space that contains more correct fixes.

Our evaluation on two widely-used benchmarks shows that \tool correctly fixes 57 Defects4J bugs and 26 QuixBugs bugs, outperforming all existing APR techniques on both benchmarks. 

\end{abstract}











\lin{be consistent globally. correct fixes? correct fixed lines? fixed lines?}\nan{updated, I will use correct fixes}






\todo{Is the caption size the default one? Captions are very small and hard to read}\thibaud{checked}





\lin{the differences between our baseline (FConv) vs coconut, additional smaller contributions: sharing vocabulary between buggy line and fix lines, embedding sharing? readers may compare coconut paper numbers with our baseline}\thibaud{Explained in RQ2}

\section{Introduction}
\noindent Automatic program repair is crucial to reduce manual software debugging efforts~\cite{le2012genprog,long2015staged,saha2017elixir,ocariza2014vejovis,xuan2017nopol,martinez2016astor,long2017automatic,xin2017leveraging,xiong2017precise,jiang2018shaping,hua2018sketchfix,wen2018context,liu2018mining,chen2017contract,le2016history,kim2013automatic,wang2016automatically,Li2020dlfix,lutellier2020coconut,chen2018sequencer,tufano2019empirical,santos2017finding,gupta2017deepfix,mesbah2019deepdelta}. There has been recent adoption of neural machine translation, a widely used technique for natural language processing (NLP) tasks, to generate correct code automatically given buggy source code~\cite{lutellier2020coconut,tufano2019empirical,chen2018sequencer,tufano2018empirical,santos2017finding,gupta2017deepfix,mesbah2019deepdelta, Li2020dlfix}. Thanks to the strong learning capabilities of NMT models, NMT-based APR techniques have outperformed most existing rule-based approaches~\cite{lutellier2020coconut,chen2018sequencer,Li2020dlfix}. NMT models use deep-learning techniques to encode buggy source code as intermediate representation in the latent space automatically, and then decode the encoded representation into target correct code. By minimizing the loss function and updating the weight parameters, NMT models learn to capture the hidden relationship between buggy code and correct code without any manual design of fix patterns or feature templates.

For a search-based APR approach (including NMT-based techniques) to generate a correct fix, it needs to satisfy two conditions: 
(1) the correct fix must be in the \emph{search space}, which is the set of all patches that the APR approach can generate, and 
(2) the \emph{search strategy} must be effective to find the correct fix in a reasonable amount of time.
Given that a correct patch is in the search space, it is desirable that the search space is small, so that it is easier to find the correct patch~\cite{long2016analysis}. 
Despite being among the most effective APR approaches, NMT-based approaches still fail to fix many bugs~\cite{lutellier2020coconut,chen2018sequencer,Li2020dlfix}. 

Compared to natural language text, source code has its own characteristics such as a strict syntax, code semantics, and an infinite number of possible identifiers. These characteristics impose unique challenges for NMT models to fix bugs automatically. 


\begin{figure}
    \centering
    \includegraphics[width=0.5\textwidth]{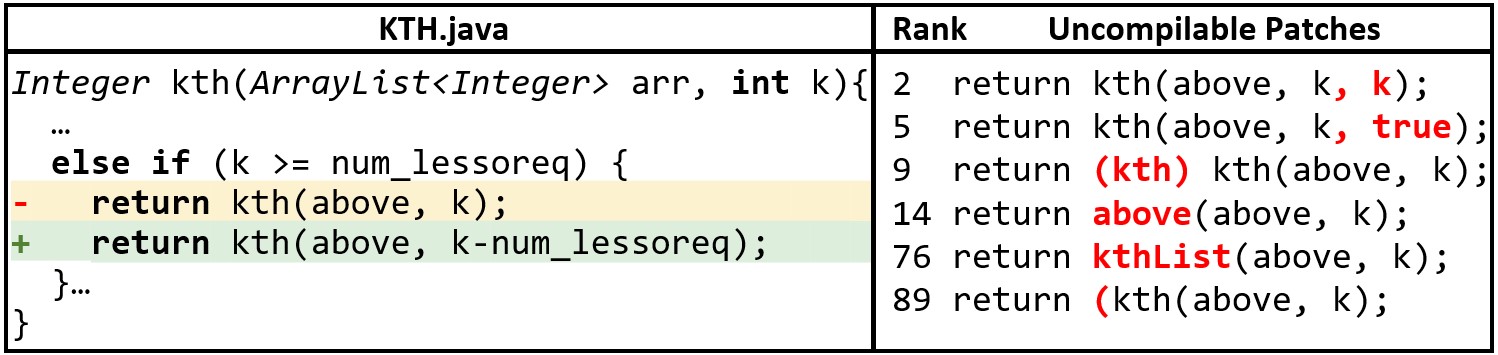}
    \caption{Uncompilable patches generated by NMT-based models, and their ranks, for a bug in QuixBugs. The line in yellow background (starting with `\texttt{-}') is the buggy line. The line in green  background (starting with `\texttt{+}') is the correct fix. The red code in generated patches disobeys Java syntax.
    }
    \label{fig:uncompilable}
\end{figure}

First, the strict syntax of source code is hard for NMT models to learn.
A major reason is that existing  techniques~\cite{lutellier2020coconut,chen2018sequencer,Li2020dlfix,ding2020patching} learn from buggy code snippets and the corresponding fixed correct code snippets (typically a few lines to tens of lines per bug), and do not use the entire source code repositories (typically millions of lines of code per project). Thus, existing NMT-based APR approaches have limited knowledge about the rigorous syntax of programming languages and the big picture of how developers write code.  
The missed opportunities are twofold: (1) existing techniques fail to take advantage of the large amount of available source code, and (2) they see partial code snippets only (which alone are often syntactically incorrect), and miss the big picture of complete methods, classes, and projects. For example, for the fix of replacing \code{while (x) \{} with \code{while (y) \{}, the open bracket \code{\{} is syntactically incorrect in this code snippet, i.e., missing the closing bracket \code{\}}. 

Such ineffectiveness is evident as demonstrated by data.  For example, up to 67\%--97\% of patches generated by the state-of-the-art NMT-based APR models~\cite{lutellier2020coconut, chen2018sequencer} are uncompilable, wasting valuable resources on incorrect patches. Figure~\ref{fig:uncompilable} shows a bug in QuixBugs and some of the top-ranked patches generated by \coconut~\cite{lutellier2020coconut}. All of these patches are uncompilable, because they call methods with wrong parameters, invoke undeclared variables, or contain mismatched parenthesis.
One important reason that \coconut fails to generate a correct patch for this bug despite generating 20,000 patches, is the large number of uncompilable patches.
The code-aware NMT-based approach we propose automatically generates a correct patch (identical to the \texttt{+} line highlighted in green) for this bug. The ranks of these uncompilable patches are high because existing NMT-based APR techniques focus on translating buggy code snippets to correct code snippets, which are partial code segments instead of full methods or programs. Since they fail to see the whole picture of the entire program or programming languages, they generate many patches with  syntax errors. 

Failing to learn how developers write code, existing NMT-based APR techniques also generate compilable but obviously-incorrect patches, as they do not look like developer-written code. These uncompilable and compilable-but-incorrect patches decrease the accuracy and efficiency of APR models, preventing APR models from generating more correct patches faster.

Second, the infinite number of possible identifiers causes NMT techniques for code to handle an enormous vocabulary if using word-level tokenization, 
where a \emph{vocabulary} contains all the unique tokens that an NMT model recognizes. Considering the complexity of NMT architectures, it is computationally too expensive for NMT-based APR models to use an enormous vocabulary. Yet with a limited vocabulary size, their search spaces do not contain all correct fixes. SequenceR~\cite{chen2018sequencer} uses a small vocabulary and shirks this complexity to a later reconstruction stage, while \coconut~\cite{lutellier2020coconut} uses a vocabulary of more than 130,000 tokens but still suffers from the \emph{out-of-vocabulary} (\emph{OOV}, i.e., an NMT model cannot recognize or generate a token) problem, resulting in its search space that still misses correct fixes.

\subsection{Our approach}
\label{sec:intro-approach}
\noindent
Thus, we propose an NMT-based approach that is specially designed to parse, analyze, model, and search source code (as opposed to natural language text) to  fix bugs automatically. Our approach, \tool, not only improves the search space (a smaller search space containing more correct patches) but also uses a more effective search strategy to find and rank correct patches higher, which are achieved through the following three main techniques that we design and use: 

\smallskip\noindent
\textbf{(1) Programming language models:} 
To help NMT models learn \devlike source code (i.e., not only compilable but also similar to those written by programmers), we apply the pre-training and fine-tuning workflow to the APR task.
Specifically, pre-trained language models have brought great improvement to many NLP tasks~\cite{clinchant2019on,skorokhodov2018semi}. They learn the probability distribution over sequences of words from a large amount of natural language text. Then one fine-tunes the pre-trained language model for a specific task by adding an extra model to it (e.g., adding a classifier for classification tasks).
The language model provides vectorized representations of input sequences to the model added to it. Since a pre-trained language model is typically trained on a larger dataset 
(since it is unsupervised learning and does not require ground truth), it offers the added model more information regarding sentence structures (e.g., syntax) and about what human-like text are (e.g., readability), which improves the 
quality of the generated text
of the fine-tuned model for the specific task significantly.

Given the effectiveness of language models in the NLP domain, we propose to add a language model pre-trained on software code, referred to as \emph{programming language (PL) model}, to an NMT architecture to create a new APR architecture. The PL model is trained to predict the next tokens in code sequences and learns to generate \devlike code. 
Then, we combine the PL model and the NMT model to form the full \emph{APR model} and fine-tune it for APR tasks. 

Our PL-enhanced NMT approach ranks correct patches higher in the search space to fix more bugs (Section~\ref{rq-contribution:gpt}). 

\smallskip\noindent
\textbf{(2) Code-aware search strategy:} When using an NMT model to generate a sequence of tokens to form a patch, ideally, one prefers the sequence with the highest score, e.g., average log probability of every token in sequence. Since this is prohibitively expensive~\cite{sutskever2014sequence}, in practice, one uses a search strategy to choose proper tokens at each step. \emph{Beam search} is a common search strategy for NMT that keeps the most $n$ probable sequences at each step, where $n$ is the \emph{beam size}.

The beam size of NLP tasks is typically 5 to 20~\cite{felix2019neural,sutskever2014sequence}. Since source code has more possible identifiers and a bigger search space than natural languages~\cite{chen2018sequencer,lutellier2020coconut}, the NMT models for APR usually require larger beam sizes (100~\cite{chen2018sequencer} to 1,000~\cite{lutellier2020coconut}) to generate enough candidate patches. However, with large beam sizes, the vanilla beam search chooses many bad patches, either uncompilable or far from correct in length. 

To address this challenge, we propose two solutions: valid-identifier-check strategy and length-control strategy. First, since source code is a formal language, only valid tokens are allowed, including  keywords and variables in scope. Invalid tokens make a patched program uncompilable, let alone capable of passing test cases. Therefore, we propose and design a \textbf{valid-identifier-check strategy} to improve the vanilla beam search, which performs static analysis to identify all valid identifiers and then forces beam search to generate only sequences with valid tokens.

Second, with a large beam size, beam search finds many very short sequences such as ``\texttt{\{}'' and ``\texttt{try \{}'', which are incorrect code snippets to fix bugs. Since correct fixes in our training data are typically of similar length to the buggy lines, we use a \textbf{length-control strategy} to punish too-short and too-long sequences to prompt \tool to generate patches of a similar length to the buggy line. 

Our code-aware beam-search strategy finds more correct fixes by generating more compilable patches and patches of similar length to the buggy lines. (Section~\ref{rq-contribution:beamsearch}).

\smallskip \noindent \textbf{(3) Subword tokenization:} The enhanced word-level tokenization proposed by CoCoNuT~\cite{lutellier2020coconut} reduces the vocabulary size of code, by using camel letters, underscores, and numbers to split long identifiers. However, many compound words (such as ``\texttt{binsearch}'' for binary search) do not contain these special characters. The previous parsing approach keeps ``\texttt{binsearch}'' as one word, which is OOV, instead of breaking it into ``\texttt{bin}'' and ``\texttt{search}'', both of which are in the vocabulary. Thus, we use \emph{byte-pair encoding} (BPE), a type of subword tokenization techniques, to tokenize compound words and rare words to further address the OOV problem. 

BPE improves the search space by both including more correct patches and reducing its size (Section~\ref{rq-contribution:subword}). \nan{changed the order}

\begin{figure*}[t]
    \centering
    \includegraphics[width=0.8\textwidth]{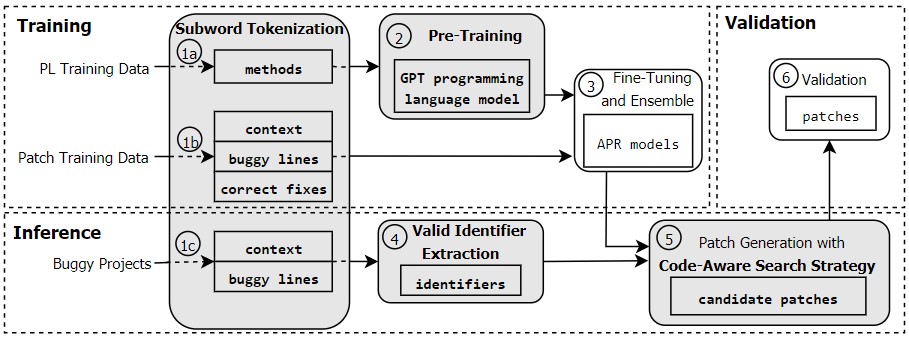}
    \caption{Overview of \tool. Grey boxes represent major novelties of \tool. Circled numbers indicate the steps of generating patches with \tool.}
    \label{fig:overview}
\end{figure*}

\subsection{Contributions}
\noindent We design and implement a code-aware NMT-based technique, \emph{\tool}, to  fix bugs automatically. Our contributions include:
\begin{itemize}
    \item An approach to 
    pre-train a PL model for APR on a very large software codebase---4.04 million methods from 1,700 open-source projects---to capture code syntax and \devlike source code, 
    \item A new APR architecture that combines a pre-trained PL model and NMT architectures to learn both code syntax and fix patterns,
    \item A new code-aware beam-search strategy, which uses valid-identifier-check and length-control strategies to find more correct fixes,
    \item A new application of subword tokenization to the APR task, which addresses the OOV problem effectively, and 
    \item A new APR approach, \tool, that combines the techniques above, and its evaluation on two  widely-used benchmarks---Defects4J and QuixBugs, where \tool fixes the most number of bugs, 57 and 26 bugs respectively, outperforming all existing APR tools. CURE is the first NMT-based approach that outperforms all state-of-the-art APR approaches on Defects4J. 
\end{itemize}

\smallskip
\smallskip
\noindent\textbf{Availability:}
Data is available in a GitHub repository\footnote{\url{https://github.com/lin-tan/CURE}}. 

\section{Background}
\label{sec:background}

\smallskip
\noindent \textbf{Candidate, Plausible and Correct Patches:} Following previous work~\cite{lutellier2020coconut}, we call patches generated by the models candidate patches. Patches that pass the validation stage are plausible, and patches identical or semantically equivalent to developers' patches are called correct patches.

\smallskip
\noindent \textbf{Parameters and Hyperparameters:}
Parameters are the weights between the connections of the network. These parameters are optimized during the training phase. Hyperparameters are arguments of the network defined before the training process. They generally include layer dimensions, number of layers, and optimization parameters.

\smallskip
\noindent \textbf{Pre-Training and Fine-Tuning:}
Pre-training is the process of training a model for a general task (e.g., next word prediction) with a very large dataset. After pre-training, one gets a pre-trained model with updated parameters. A pre-trained model can be fine-tuned for a similar but specific task (e.g., text generation) with few training data. During fine-tuning, usually one needs to add extra models to the pre-trained model to fit the task, and the parameters of both the pre-trained model and added models are updated.

\smallskip
\noindent \textbf{Context-aware neural machine translation (CoNuT) architecture:}
We use \cocoarchitecture as our NMT architecture in this paper. \cocoarchitecture consists of a buggy lines encoder, a context encoder, a merger, a decoder, an attention module, and a token generation module, where the encoders and decoder are implemented with convolutional sequence-to-sequence  architecture~\cite{gehring2017convolution}. The details of \cocoarchitecture is described in~\cite{lutellier2020coconut}. \cocoarchitecture has shown good results for APR, and convolutional architecture can be stacked to capture hierarchical features and long dependencies for larger contexts~\cite{lutellier2020coconut}.
\section{Approach}

\noindent
To address the challenges described in the Introduction, we design and apply three novel techniques, i.e., \emph{subword tokenization} to improve the search space (Section~\ref{sec:tokenization}), a \emph{programming language model}\lin{Did we ever spell out and introduct what GPT is? we should}\nan{move GPT to PL model subsection} to learn \devlike source code and improve patch ranking (Section~\ref{sec:gpt} and Section~\ref{sec:tuning}), and a new \emph{code-aware beam-search strategy} (Section~\ref{sec:search}) to improve patch ranking and  generate more correct patches.

\subsection{Overview}
\smallskip \noindent Figure~\ref{fig:overview} presents an overview of our approach. \tool consists of three stages: training, inference, and validation. During the training stage, \tool extracts methods from open-source projects, referred to as \emph{PL training data}, and tokenizes them (step~{\large\textcircled{\small{1a}}} in Figure~\ref{fig:overview}). Different from previous work~\cite{lutellier2020coconut,tufano2019empirical,chen2018sequencer,santos2017finding,gupta2017deepfix,mesbah2019deepdelta,Li2020dlfix}, we use \textbf{subword tokenization}, which produces a smaller but more accurate search space that contains more correct patches. \tool uses these tokenized methods to train a \textbf{new programming language model}\thibaud{Either don't say GPT or spell it out since it's the first occurence of GPT}\nan{I removed GPT. I think introduce both GPT/CoNuT in later detailed section is more consistent} that learns \devlike source code with correct syntax (step~{\large\textcircled{\small{2}}}). \tool also tokenizes the buggy lines, context, and correct fixes extracted from the commit history of open-source projects, referred to as \emph{patch training data}, into sequences of tokens (step~{\large\textcircled{\small{1b}}}). We use these sequences to fine-tune an ensemble of $k$ APR models (step~{\large\textcircled{\small{3}}}). Each APR model combines the PL model with a context-aware neural machine translation (\cocoarchitecture) model ~\cite{lutellier2020coconut}. 

During the inference stage, a user provides a buggy project along with the location of buggy lines to \tool. These are standard input that existing APR tools require~\cite{chen2018sequencer,liu2019tbar,xuan2017nopol,le2012genprog,qi2013does,lutellier2020coconut}. \tool tokenizes the buggy and the context lines (step~{\large\textcircled{\small{1c}}}), then analyzes the source code to extract a list of \emph{valid identifiers} that are in scope of the buggy lines (step~{\large\textcircled{\small{4}}}). The patch generation module generates a list of candidate patches using a \textbf{new code-aware beam-search strategy} (step~{\large\textcircled{\small{5}}}). This new algorithm discards many irrelevant patches on the fly (i.e., as soon as an invalid token is generated) and penalizes patches that are unlikely to be correct (e.g., fixes that are very different from the buggy line in length), which saves a lot of resources and allows \tool to search deeper for correct patches. 

In the validation stage, \tool validates candidate patches by compiling and executing the test suites of the patched projects. \tool outputs a list of plausible patches (step~{\large\textcircled{\small{6}}}) for developers to examine. 

\smallskip
\subsection{Data Extraction}
\label{sec:extraction}
\noindent \tool uses two different types of training data. First, the GPT PL model is trained on millions of methods extracted from open-source Java projects. 
Second, \tool fine-tunes the PL model for the APR task. This step requires APR specific training data (i.e., buggy lines, context, and correct fixes). We use \coconut's training data shared on GitHub~\cite{lutellier2020coconut}. \coconut's authors extracted this dataset from open-source repositories and identified buggy commits based on keywords in commit messages (“fix,” “bug,” and “patch”). They also cleaned the dataset using commit message anti-patterns (“rename,” “clean up,” “refactor,” “merge,” “misspelling,” and “compiler warning"). Similar to \coconut, we use the method surrounding the buggy lines as context.

\subsection{Code Representation and Tokenization}
\label{sec:tokenization}
\noindent\textbf{Word-level tokenization:} To tokenize buggy-, context-, and fixed-lines to token sequences, \tool first uses enhanced word-level tokenization~\cite{lutellier2020coconut} to separate code lines by spaces, camel letters, underscores, strings, and numbers (except 0 and 1).

\smallskip \noindent\textbf{Out-of-vocabulary Issue:}
The vocabulary size after the word-level tokenization is
larger than what is commonly used in NLP and the test set still contains 2\% of OOV tokens. Excluding rare tokens is problematic for source code because OOV tokens are likely to be important project-specific tokens. Excluding such tokens makes it difficult for NMT models to fix bugs in these new projects.

Some existing NMT-based APR models~\cite{lutellier2020coconut} do not generate OOV tokens, missing the opportunity to fix more bugs. SequenceR uses a special token as a placeholder for OOV tokens, and then uses a copy mechanism to reconstruct them. The copy mechanism replaces the placeholder tokens with the most likely token from the input buggy lines. However, this solution would fail to generate some patches, since it can only copy tokens appearing in the buggy lines.

\smallskip
\noindent\textbf{Subword tokenization:}
To address the OOV problem and reduce the vocabulary size, we use \emph{byte pair encoding (BPE)}, which is an unsupervised learning algorithm to find the most frequent subwords in a corpus by merging the most frequent byte pair iteratively~\cite{rico2015neural}. BPE has been used in many NLP tasks and is useful to reduce vocabulary size and mitigate the OOV problem efficiently~\cite{rico2015neural,ashish2017attention,radford2018improving}. 

\begin{figure}
    \centering
    \subfigure[Buggy line and correct fix of Closure 62.]{
        \includegraphics[width=0.48\textwidth]{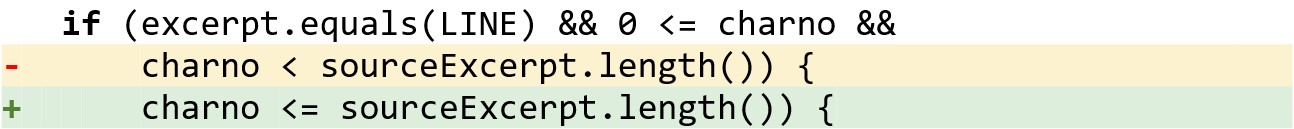}
    }
    \subfigure[Word-level tokenization result of buggy line and correct fix.]{
        \includegraphics[width=0.48\textwidth]{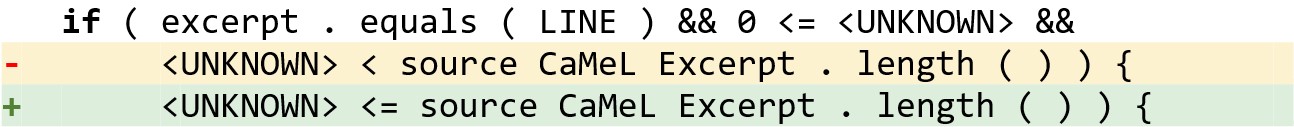}
    }
    \subfigure[Subword-level tokenization result of buggy line and correct fix.]{
        \includegraphics[width=0.48\textwidth]{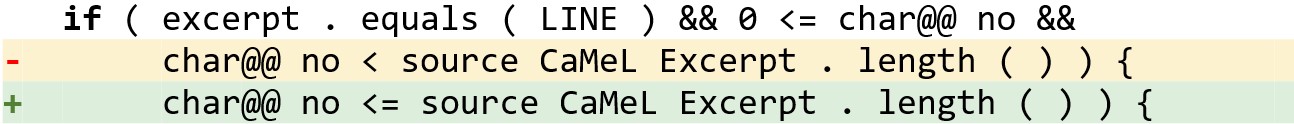}
    }
    \caption{Tokenized results that use word-level tokenization and subword tokenization of Closure 62 in Defects4J.
   }
    \label{subword-tokenization}
\end{figure}

Figure~\ref{subword-tokenization} shows an example from the inference stage demonstrating the effectiveness of the subword tokenization. Lines starting with `\texttt{-}'  are the buggy lines (input) and those starting with `\texttt{+}'  are the correct fixes. Figure~\ref{subword-tokenization}(a) shows the source code of a real bug in Defects4J~\cite{just2014defects4j}, while Figure~\ref{subword-tokenization}(b) shows the code after using the enhanced word-level tokenization. Figure~\ref{subword-tokenization}(c) shows the same code tokenized by our subword tokenization. In Figure~\ref{subword-tokenization}, each consequence separated by space is a token excluding the `\texttt{-}' and `\texttt{+}' signs. \lin{revised}\nan{checked}

When using only the enhanced word-level tokenization, the variable ``\texttt{charno}'' is an OOV token. Thus, \coconut and SequenceR fail to fix this bug since \coconut cannot generate OOV tokens and SequenceR does not fix it correctly with the copy mechanism. With our subword tokenization, ``\texttt{charno}'' is split into two tokens, both of which appear in the vocabulary---``\texttt{char@@}'' (``\texttt{@@}'' indicates that the token needs to be concatenated with the following token) and ``\texttt{no}'', enabling \tool to generate a correct patch for this bug.

By applying subword tokenization, we use a smaller vocabulary to form a smaller but better search space that contains more correct patches. Section~\ref{rq-contribution:subword} evaluates the impact of our subword tokenization approach.

\smallskip
\subsection{Programming Language Model}
\label{sec:gpt}
\noindent To address the challenges of learning \devlike source code, we train a language model on open-source programs, referred to as a \emph{programming language model (PL model)}. 
A PL model optimizes the probability of a sequence of tokens being a real-world code snippet. We use Generative Pre-trained Transformer (GPT)~\cite{radford2018improving}\nan{explained GPT here} for PL modeling because GPT has been shown to improve the performance of many different NLP tasks~\cite{radford2018improving,radford2018language}.

Pre-training a PL model allows for separating programming language learning from patch learning. The advantages are twofold. First, GPT learning is unsupervised and only requires complete methods; therefore one can extract a large amount of data automatically and accurately to train it. Second, during fine-tuning, the APR model already knows the PL syntax (thanks to the PL model), making the fine-tuning more efficient.

Given a sequence of tokens representing a method, $\mathbf{x}=(x_1,..., x_n)$, where $x_i$ is a token in the method sequence $\mathbf{x}$, the PL modeling objective is to maximize the average likelihood:
\begin{equation}
    L_{GPT}(\mathbf{x}) = \frac{1}{n}\Sigma_{i=1}^n{\log P(x_i | x_1,..., x_{i-1};\Theta)}
\end{equation}
where $\Theta$ represents matrices of trainable weights of the PL model. $P(x_i | x_1,..., x_{i-1};\Theta)$ is the conditional probability of token $x_i$ being the next token,  given a sequence of $x_1,..., x_{i-1}$, which is calculated by the PL model with weights $\Theta$. 
At a high-level, the objective of the PL model training is to find the best weights ($\Theta$) so that sequences of tokens $x_1,..., x_{n}$ representing real methods in projects obtain a higher $L_{GPT}$ score than other sequences. Since methods in popular open-source projects are dominantly well-formed correct blocks of code, we use them to train our PL model to learn if a given sequence of tokens is likely to form real-world code (compilable and looks like written by programmers).

\smallskip
\subsection{Fine-Tuning for APR with a PL Model} \label{sec:tuning}

\noindent After pre-training the PL model, \tool fine-tunes the GPT PL model for the APR task by combining it with an NMT model as the APR model. 
We use the \cocoarchitecture (Section~\ref{sec:background}) as \tool's NMT architecture. \nan{revised}

The APR model takes buggy lines and their context as input and aims to generate a correct patch. 
During the fine-tuning process, the APR model is trained to learn the transformation from the buggy lines and context (e.g., the buggy method) to the correct fix. We use \mbox{$\mathbf{x_b}=(x_{b_1},...,x_{b_n})$} to denote the buggy lines, \mbox{$\mathbf{x}=(x_1,...x_{c_n})$} to denote the context,  and \mbox{$\mathbf{y}=(y_1,...,y_{f_n})$} to denote the correct fixes, where $b_1,...,b_n$ are the indices of the buggy lines in the context, while
$c_n$ and $f_n$ are the lengths of the context and correct fixes respectively. 

We denote the weights of the PL model as $\Theta$ and weights of \cocoarchitecture as $\Phi$.
The APR model is fine-tuned by updating $\Theta$ and $\Phi$ to maximize the following average log-likelihood:
\begin{equation}
\begin{aligned}
    L_{NMT}(\mathbf{x},\mathbf{x_b},\mathbf{y}) & = \frac{1}{f_n}\Sigma_{i=1}^{f_n}\log P(y_i|\mathbf{x},\mathbf{x_b},y_0,\mathrm{\ldots},y_{i-1};\Theta,\Phi) \\
    y_0 & = x_{b_1-1} \\
\end{aligned}
\end{equation}
$P(y_i|\mathbf{x},\mathbf{x_b},y_0,\ldots,y_{i-1};\Theta,\Phi)$ is the conditional probability calculated by the APR model with weights $\Theta$ and $\Phi$, where $y_i$ is the token following the sequence $(y_0,y_1,\ldots,y_{i-1})$ in the correct fix, given the buggy lines $\mathbf{x_b}$ and context $\mathbf{x}$. 
For the first token in the correct fix, the probability is the conditional probability of token $y_1$ given $y_0$, where $y_0$ is the token right before the correct fix, i.e., $x_{b_1-1}$. For example, the entire method ``\texttt{kth()}'' is the context in Figure~\ref{fig:uncompilable}, while the buggy lines and the correct fixes start at the same index in the context, and $y_0$ is the token $\{$ right before the ``\texttt{return}'' statement. 

To prevent the PL model from losing the information it learned during pre-training, we include the language modeling (i.e., $L_{GPT}$) as an auxiliary objective to the fine-tuning process. It also improves the generalization of the fine-tuned model~\cite{radford2018improving}. Therefore, the APR model is fine-tuned by maximizing the combined average log-likelihood:

\begin{equation}
\begin{aligned}
    L_{APR}(\mathbf{x,x_b,y}) & = L_{NMT}(\mathbf{x,x_b,y}) + \lambda L_{GPT}(\mathbf{y'}) \\
    \mathbf{y'} & = (x_1,x_2,...,x_{b_1-1},\mathbf{y})
\end{aligned}
\end{equation}

\begin{figure}
    \centering
    \includegraphics[width=0.5\textwidth]{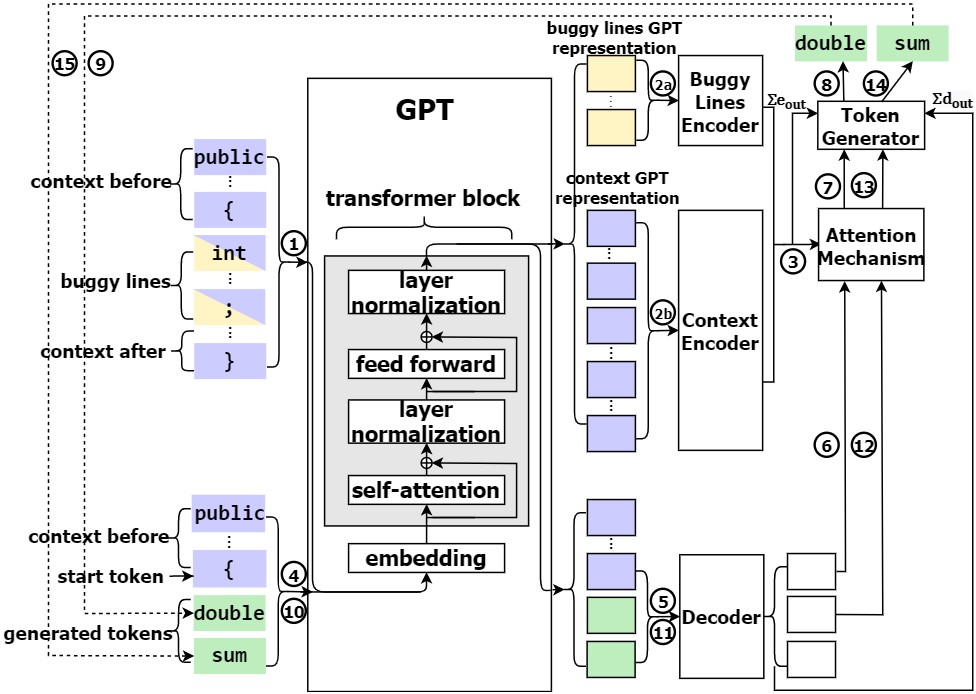}
    \caption{Architecture of the APR models used in \tool. Yellow, purple and green boxes refer to the buggy lines, context and the generated patch.} 
    \label{model architecture}
\end{figure}

\noindent where $\mathbf{y'}$ is the token sequence from the beginning of the buggy method to the last token in the correct fix ($x_1,x_2,...,x_{b_1-1}$ is the prefix of $\mathbf{x}$ before $\mathbf{x_b}$). Probability $L_{GPT}(\mathbf{y'})$ is the likelihood of $\mathbf{y'}$ being a real source code snippet, while $\lambda$ is a hyperparameter referring to the coefficient of $L_{GPT}$ in the combined log-likelihood $L_{APR}$.

The fine-tuning stage aims to find the best set of parameters $\Theta$ and $\Phi$ to maximize $L_{APR}$ for all buggy lines, context, and correct fixes in the training data.

In the training mode, the APR model takes the pre-trained GPT module (the PL model) and the patch training data as input. The patch training data consists of the buggy lines, the context, and the correct fixes. We train the APR model for multiple epochs (i.e., multiple passes on the training data) to obtain the best combination of weights $\Theta$ and $\Phi$.

In the inference mode, the APR model has  access to only the buggy lines and their context and  outputs a sequence of tokens representing the patch. Figure~\ref{model architecture} shows a simplified view of the architecture of our combined APR model and how the model is used in inference mode. Our APR model consists of two components: a PL model (GPT) and an NMT model (\cocoarchitecture). 

First, \tool generates the GPT representation of the context lines (step {\large\textcircled{\small{1}}} in Figure~\ref{model architecture}). As explained in Section~\ref{sec:gpt}, the GPT model was trained on complete methods, therefore the input of the GPT model needs to be a method. If we directly feed the first token of the buggy line to the GPT model (``\texttt{int}'' in Figure~\ref{model architecture}), the GPT model will generate a bad embedding for it since it expects the first token of a sequence to be the first token of a method (e.g., ``\texttt{public}''). 

Hence, the GPT model generates an embedding for all tokens in the buggy method. The \cocoarchitecture model contains two encoders. The buggy lines encoder only takes the representation of the buggy line as input. Therefore, \tool extracts the subsequence that corresponds to the buggy line embedding from the buggy method embedding (yellow boxes in Figure~\ref{model architecture}) and forwards it to the buggy lines encoder (step {\large\textcircled{\small{2a}}}). The second encoder is for the context and takes the embedding of the entire buggy method (purple boxes in Figure~\ref{model architecture}) as input (step {\large\textcircled{\small{2b}}}). \tool merges the output of the two encoders (step {\large\textcircled{\small{3}}}) before sending it to the attention mechanism and the token generator.

To start generating tokens, the attention mechanism and the token generator need the encoder's and the decoder's output. At the start of the inference, none of the fixed tokens have been generated yet. \coconut started the decoding sequence with an initial ``\texttt{<START>}'' token. However, it is better to initialize the sequence with the last token of the context before the buggy line to provide additional contextual information. To obtain the embedding of this token,  we pass the context before the buggy line to the GPT model (step  {\large\textcircled{\small{4}}}) and then feed the embedding of the last token (``\texttt{\{}'') to the decoder (step {\large\textcircled{\small{5}}}). The decoder generates a representation of the token, which is forwarded to the attention mechanism (step {\large\textcircled{\small{6}}}).

The attention mechanism combines the output of the two encoders and the output of the decoder to form the attention map between the last token (``\texttt{\{}'' in the example) and the buggy method. Then, the token generation outputs the first token of the fixed sequence (``\texttt{double}'' in step {\large\textcircled{\small{8}}}). This token is then appended to the decoder input (step {\large\textcircled{\small{9}}}).
Then, the decoder starts the next iteration (steps {\large\textcircled{\small{10}}} to {\large\textcircled{\small{15}}}) with the input ``\texttt{\{ double''} and generates the token ``\texttt{sum}''. This iterative process continues until the end-of-sequence token ``\texttt{<EOS>}'' is generated.  

\subsection{Ensemble Learning}
\label{sec:ensemble}
\noindent
Prior work~\cite{lutellier2020coconut} shows that ensemble learning, i.e., combining multiple models, enables NMT-based APR approaches to fix more bugs: the number of bugs correctly fixed rises from 22 to 44 when the number of models increases from 1 to 20. Therefore, we combine (1) models with different hyperparameters and (2) models with two different architectures (\cocoarchitecture and FConv~\cite{gehring2017convolution}) for our ensemble learning. 
The GPT PL model is general as it represents the entire PL. Thus, each APR model starts with the same PL model, fine-tunes it, and combines it with \cocoarchitecture or FConv architectures that have different hyperparameters (step {\large\textcircled{\small{3}}} of Figure~\ref{fig:overview}).

Balancing the computation cost and tuning effectiveness, we use random search to pick different hyperparameter values (e.g., number of convolution layers, convolution dimensions, and dropout) in a reasonable range and tune each model for one epoch. Based on  each model's perplexity (i.e., a measurement of how well a model predicts an instance) on our validation data, we choose the top $k$  models for ensemble learning and keep training them until convergence.

\subsection{Code-Aware Beam-Search Strategy and Patch Generation} 
\label{sec:search}
\noindent The goal of patch generation is to generate the sequence with the highest probability given the buggy line and its context. The APR model generates one token with its probability at a time. Searching for the sequence with the highest probability is  exponential in the length of the output sequence. Thus, we need an effective search strategy to find a sequence with a high probability.

Beam search is an optimized greedy strategy and the most common search strategy used for NMT. Beam search keeps only the $n$ ($n$ is beam size, a hyperparameter of beam search) optimal nodes, instead of all nodes, in the search tree to expand at every step and remove the rest. A major issue of the vanilla beam search is that it considers only the log-probability provided by the model to generate the next token. Since other information about code (e.g., variables in scope) is unavailable to the APR model, it often generates a high score for an out-of-scope variable, producing an uncompilable candidate patch.

Therefore, we design two techniques---\emph{valid-identifier check} and \emph{length control}---to make the beam search code-aware.

\begin{figure}[t]
    \centering
    \includegraphics[width=0.49\textwidth]{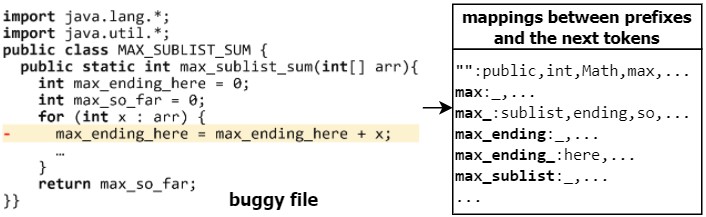}
    \caption{An example of extracting mappings between prefixes and valid next tokens from buggy projects. Line with yellow background is buggy line.}
    \label{analyze identifier}
\end{figure}

\smallskip
\noindent \textbf{Valid-Identifier Check:} 
Only a few tokens are valid in a certain Java code snippet since correct code must follow Java syntax and compilation rules. To generate valid identifiers only, \tool first uses static analysis to analyze and extract valid identifiers. Then \tool tokenizes these identifiers (e.g., ``\texttt{max\_ending\_here}'' becomes $[$\texttt{max}, \texttt{\_}, \texttt{ending}, \texttt{\_}, \texttt{here}$]$), and builds the mappings between all prefixes and their valid succeeding tokens (as showns in Figure~\ref{analyze identifier}). These mappings are necessary for the beam-search algorithm to know that after generating the sequence ``\texttt{max\_ending\_}'', ``\texttt{here}'' is a valid next token because ``\texttt{max\_ending\_here}'' is a valid identifier.

At every decoding step, the NMT model outputs a probability distribution of all the tokens in vocabulary. \tool's new valid-identifier-check strategy first analyzes the token sequence already generated to get the ``prefix'' of the next token needed to be generated. If the generated token sequence does not contain any prefix, (i.e., the next token will be the beginning of a new identifier), the ``prefix'' will be set to the empty string. Then, based on the mappings between all possible prefixes and their valid succeeding tokens, the valid-identifier-check strategy modifies the probability distribution and sets the probability of invalid tokens to $-inf$. By doing this, the valid-identifier-check strategy discards many impossible nodes, increasing the possibility of finding the correct patch.

\begin{figure}[t]
    \centering
    \subfigure[Vanilla beam search vs. beam search using valid-identifier check, with beam size of \textbf{2}]{
        \includegraphics[width=0.48\textwidth]{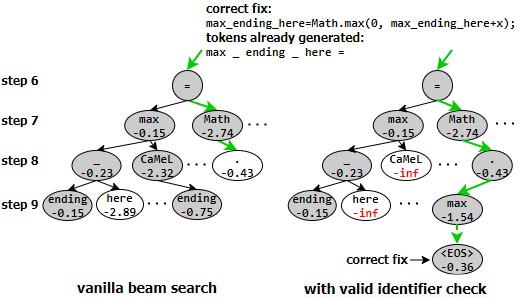}
    }
    \subfigure[Vanilla beam search vs. beam search using length control in the same bug, but with beam size of \textbf{1,000}]{
        \includegraphics[width=0.48\textwidth]{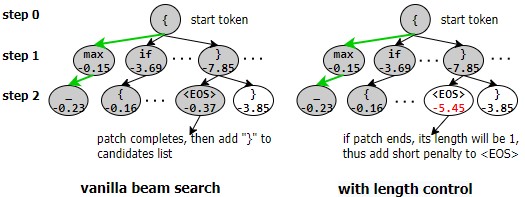}
    }
    \caption{Examples using the vanilla beam search and beam search with valid-identifier-check and length-control strategies. \textcolor[RGB]{0,175,0}{\textbf{\emph{Green arrows}}} are the paths to the correct fixes. \colorbox{light-gray}{\textbf{\emph{Grey circles}}} are the nodes kept by the search strategies in the search tree at every level, and \emph{white circles} are nodes discarded. \textcolor{red}{\textbf{\emph{Red numbers}}} are the log probability changed by search strategies. }
    \label{beam search}
\end{figure}


Figure~\ref{beam search}(a) illustrates how our code-aware beam search outperforms the vanilla beam search. The correct fix is {\small``\texttt{max\_ending\_here=Math.max(0,max\_ending\_here)}''}, and the start of the output sequence (``\texttt{max\_ending\_here=}'') has already been generated in steps 1 to 6. We use a beam size of 2 to simplify the illustration. The path to the correct fix is marked with green arrows and the nodes considered by the beam-search strategies are in light grey.

During step 7, the two most likely nodes according to the APR model are ``\texttt{... max}'' and ``\texttt{... Math}'', where ``\texttt{...}'' refers to ``\texttt{max\_ending\_here=}'' which has  already been generated. However, at step 8, the average log-likelihood of ``\texttt{... Math .}'', which is the sequence denoted by the green path in the left subfigure of Figure~\ref{beam search}(a), is less than that of ``\texttt{... max \_}'' and ``\texttt{... max CaMeL}'', so the vanilla beam search drops it. Thus, the entire subtree containing the correct fix is excluded. 

In contrast, with our valid-identifier-check strategy, the average log-likelihood of ``\texttt{... max CaMeL}'' is set to \texttt{-inf} because there is no valid identifier starting with ``\texttt{max CaMeL}''. Therefore, our code-aware beam search keeps searching the subtree of node ``\texttt{... Math}'', which leads to the correct fix.

\smallskip
\noindent \textbf{Length Control}: In our training data, most correct fixes have a similar length as the buggy lines. We find the length difference of 75\% of the bugs in our 2.7 million patch training data is less or equal to 5 tokens. This means that most of the time, the correct fixes are small modifications to the buggy lines, and more complex changes are less common.

Therefore, we use length-control strategy to generate patches of a similar length of the buggy lines, by punishing short and long patches. At every decoding step, the length-control strategy calculates the length of the sequence already generated. If the current length is much smaller than the buggy lines, it decreases the log-likelihood of ``\texttt{<EOS>}'' to prevent this patch from reaching the end. And if the current length is already much larger than the  buggy lines, it increases the log-likelihood of ``\texttt{<EOS>}'' to prompt this patch to end.

To determine the penalty value, we leverage the length difference distribution in the patch training data to calculate the log-probability of each length difference, denoted as function $F_{len}$. Our length-control strategy modifies the log-likelihood of token ``\texttt{<EOS>}'' by adding the following penalty to it:

$$penalty =
\begin{cases}
0 & -5 <= l_b-l_p <= 5  \\
F_{len}(l_b-l_p) & \text{otherwise}
\end{cases}$$

\noindent where the lengths of the buggy lines and  the patch sequence already generated are $l_b$ and $l_p$ respectively. We empirically set a tolerance threshold of 5 to increase flexibility.

Figure~\ref{beam search}(b) illustrates this issue. The bug is the same as in Figure~\ref{beam search}(a) but with a larger beam size of 1,000. In step 2, the sequence ``\texttt{\{ \}}'' reaches the ``\texttt{<EOS>}'' token. Using the vanilla beam search, patch ``\texttt{\{ \}}'' has a low average log-probability but is still in the top 1,000, this is added to candidate patches because the beam size is large (1,000). Such low score patches take up the valuable slots and prevent correct patches from being selected. In our code-aware beam-search strategy, since the complete sequence, ``\texttt{\{ \}}'', is much shorter than the buggy sequence, the ``\texttt{<EOS>}'' token receives a large penalty and is not selected as a candidate node (not included in the 1,000 highest average log-probabilities), allowing the search strategy to search deeper along other paths.

While \tool focuses on fixing bugs whose fixes are similar in length to the buggy lines, our length-control strategy is general and can be adapted to generate longer patches by modifying the penalty weights. Given the complexity of APR, fixing similar-length bugs and other bugs separately may be an effective way to decompose this complex task.

\subsection{Patch Validation}
After APR models generate candidate patches, we reconstruct the token sequences to code statements. We first concatenate tokens end with ``\texttt{@@}'' to their successors, which is the reconstruction from subwords to words. Then we extract donor code from the buggy code file to reconstruct the abstracted tokens (numbers and strings).

Reconstructed statements are ranked by the average log-probability of their tokens and then inserted into the buggy code file to replace the buggy lines. Every patched project is compiled to filter the uncompilable patches and we run the test suites until we find a patch that satisfies two conditions: (1) passing all the test cases that the buggy project passes and (2) passing at least one test case failed on the buggy project, which are the same criteria for validation used in previous work~\cite{lutellier2020coconut,yang2017better}. \lin{is it all test cases or passes all test cases that the buggy code passes and the failure exposing tests following \coconut, as Opad shows that this metric is more robust}\nan{revised}

\section{Experimental Setup}

\noindent \textbf{Realistic Evaluation:}
To make the evaluation realistic, we need to avoid using future data~\cite{lutellier2020coconut,tan2015online}. We address this issue by using data committed before the first bug in our benchmark (i.e., 2006) for pre-training, fine-tuning, and validation.

\smallskip
\noindent \textbf{PL Training Data:} We download all (1,700) open-source Java projects from GitHub that have at least one commit before the first bug in Defects4J according to GHTorrent~\cite{gousios2012ghtorrent} and roll them back to a version before 2006 to avoid using future data. Then, we use JavaParser~\cite{smith2019javaparser} to extract all methods except abstract methods and those longer than 1,024 tokens.
The PL training data contains 4.04 million methods, with 30,000 instances for validation.

\smallskip
\noindent \textbf{Patch Training Data:} 
We use \coconut's training data shared on GitHub~\cite{lutellier2020coconut} as our patch training data, which is extracted from 45,180 Java projects. These Java projects are a superset of the projects used for PL training data since we need more projects to extract enough patch data and it is too expensive to use all these projects for PL training. Then we discard the instances whose context or correct fixes are longer than 1,024 tokens after subword tokenization. Removing instances from the training set is a common practice for machine learning, and since the test set (bugs in Defects4j and QuixBugs are untouched), this setup is valid. Our patch training data contains 2.72 million training instances and 16,920 validation instances.

\smallskip
\noindent \textbf{Subword Tokenization, Training, Fine-Tuning, and Inference:} 
We set the target vocabulary size to be 50,000 for BPE. For the GPT model, considering previous work's recommendation~\cite{radford2018improving} and our hardware limits, we use an embedding size of 384, eight layers of transformer blocks, and six attention heads. We train GPT for five epochs, using a batch size of 12. We use Adam optimizer~\cite{kingma2014adam}, and the learning rate increases from 0 to $2.5e^{-4}$ at the first 2,000 training steps and then decreases using a cosine schedule.
 
To fine-tune the hyperparameters of an APR model, we use random search with the following ranges: convolution dimension (128--512), kernel size (2--10), number of convolutional layers (1--5), and dropout (0--0.5). $\lambda$ is empirically set to 0.3. We train 100 APR models on a smaller subset of patch training data for one epoch and keep the top five models combining GPT with \cocoarchitecture model and top five APR models combining GPT with FConv model. We use Adam optimizer with a learning rate of $6.25e^{-5}$ to keep tuning the top models on our patch training data for one epoch, with a batch size of 12. 

In inference mode, we use beam search with a beam size of 1,000, and \tool generates 10,000 candidate patches for every bug. During the validation stage, considering the time cost and that most correct fixes have high ranks, we validate the top 5,000 candidate patches per bug. 

\smallskip
\noindent \textbf{Infrastructure:} We use GPT implemented by Hugging Face ~\cite{wolf2019huggingface}, \cocoarchitecture and FConv implemented using fairseq ~\cite{ott2019fairseq,lutellier2020coconut}. We train and evaluate our models on one 56-core server with one NVIDIA TITAN V and three Xp GPUs.

\section{Evaluation and Results}

\noindent We use two widely-used benchmarks, Defects4J (v1.4.0)~\cite{just2014defects4j} and QuixBugs~\cite{lin2017quixbugs} for evaluation. \todo{make sure we state the Defects4J version we use, esp. if Defects4J updated its bug numbering system,}\nan{added the defects4j version} Following~\cite{lutellier2020coconut}, we remove two Defects4J bugs, Closure 63 and Closure 93, from our evaluation as they are duplicates of other Defects4J bugs. We compile the patched projects and run the test suites to find plausible patches, i.e., patches that pass the relevant test cases\thibaud{Should be consistent with what we say in approach. It should be "relevant test cases"}\nan{revised}. Two co-authors independently  check  plausible patches and consider correct only those that are identical or semantically equivalent to  developers' patches (92\% of agreement, Cohen's k of 0.84), then discuss to resolve disagreements.

We compare \tool with 25 APR techniques~\cite{lutellier2020coconut,Li2020dlfix,chen2018sequencer,liu2019tbar,saha2019harnessing,qi2013does,xuan2017nopol,liu2018mining,saha2017elixir,chen2017contract,wen2018context,xin2017leveraging,hua2018sketchfix,qi2015analysis,le2012genprog,yuan2018arja,xiong2017precise,martinez2018ultra,durieux2016dynamoth,jiang2018shaping,liu2019you,liu2019avatar,koyuncu2018fixminer,le2016history,liu2018lsrepair}.  
Table~\ref{evaluation} shows the comparison results. 
The table lists only a few top-ranked techniques in terms of the number of bugs that they fix in each benchmark, including state-of-the-art pattern-based techniques~\cite{liu2019tbar,saha2019harnessing}, three NMT-based techniques~\cite{lutellier2020coconut,chen2018sequencer,Li2020dlfix} and the techniques that have been evaluated on QuixBugs. None of these tools uses subword tokenization, pre-trained PL model or, code-aware search strategy. Other tools (e.g., AVATAR~\cite{liu2019avatar}, kPAR~\cite{liu2019you}, SimFix~\cite{jiang2018shaping}) either fix fewer bugs than the listed tools or were not evaluated on these benchmarks.
Results from \emph{all} 9 techniques in Table~\ref{evaluation} except Astor~\cite{martinez2016astor} and Hercules~\cite{saha2019harnessing} use the perfect fault localization (FL) of bugs to report  bug fixing results. As stated in previous work~\cite{liu2019you,liu2019avatar}, having APR techniques use the same FL techniques (e.g., perfect FL) is a fair way to compare APR techniques since different FL methods affect APR techniques differently.
\tool's correctly generated patches are available in our GitHub Repository. 

\newcolumntype{P}[1]{>{\centering\arraybackslash}p{#1}}
\begin{table}[]
    \caption{Comparison with state-of-the-art APR tools. The numbers are shown as x/y, where x is the number of bugs fixed correctly and y is the number of bugs with plausible patches. `-' means the approach has not been evaluated on that benchmark.}
    \centering
    \begin{tabular}{p{4cm}P{1.6cm}P{1.6cm}}
    \hline
        \textbf{Tool} & \textbf{Defects4J} & \textbf{QuixBugs} \\
         & 393 bugs & 40 bugs \\
    \hline
        \textbf{Astor}\cite{martinez2016astor} & - & 6/11 \\
        \textbf{Nopol}\cite{xuan2017nopol} & 2/9 & 1/4 \\
        \textbf{RsRepair}\cite{qi2013does} & 10/24 & 2/4 \\
        \textbf{Hercules}\cite{saha2019harnessing} & 49/72 & - \\
        \textbf{TBar}\cite{liu2019tbar} & 52/85 & - \\
        \textbf{SequenceR}\cite{chen2018sequencer} & 14/19 & - \\
        \textbf{DLFix}\cite{Li2020dlfix} & 36/65 & - \\
        \textbf{\coconut}\cite{lutellier2020coconut} & 44/85 & 13/20 \\
    \hline
        \textbf{\tool} & \textbf{57/104} & \textbf{26/35} \\
    \hline
    \end{tabular}
    \label{evaluation}
\end{table}

\subsection{RQ1: How does \tool perform against state-of-the-art APR techniques?}
\noindent
In Table~\ref{evaluation}, the results are displayed as x/y, where x is the number of bugs fixed correctly and y is the number of bugs with plausible patches. 

\tool fixes the most number of bugs, 57 and 26 respectively, in both Defects4J and QuixBugs. Specifically, \tool generates plausible patches for 104 Defects4J bugs, 57 of which are correctly fixed by \tool,  outperforming the best existing approach TBar by five bugs.\lin{for numbers $<=$ 10, spell it out, fix globally}\nan{fixed} Compared to NMT-based approaches~\cite{chen2018sequencer,lutellier2020coconut,Li2020dlfix}, \tool correctly fixes 13 more bugs than the best NMT-based approach \coconut. 
\tool fixes 26 QuixBugs bugs (twice as many bugs as \coconut), including 12 bugs that none of the four existing tools that have been evaluated on QuixBugs can fix. In Defects4J, \tool fixes one unique bug, Chart 17, that has not been fixed by any of the 25 existing approaches. 

\begin{figure}[t!]
    \centering
    \includegraphics[width=0.48\textwidth]{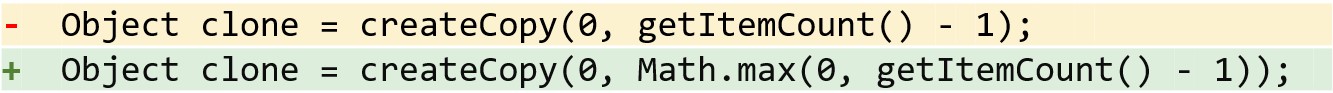}
    \caption{Chart 17 in Defects4J is a bug only fixed by \tool}
    \label{unique-all}
\end{figure}

\begin{figure}[t!]
    \centering
    \includegraphics[width=0.48\textwidth]{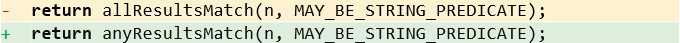}
    \caption{Closure 10 in Defects4J is a bug that \tool fixes but \coconut does not.}
    \label{unique-nmt}
\end{figure}

\begin{figure}[t]
    \centering
    \includegraphics[width=0.48\textwidth]{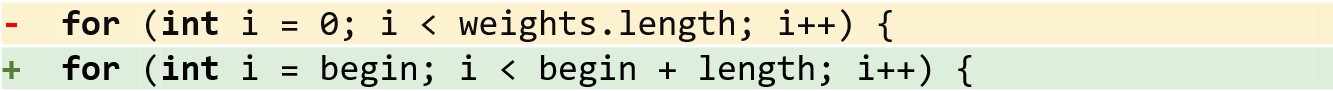}
    \caption{Math 41 in Defects4J is a bug that \tool  fixes but pattern-based tools TBar and Hercules do not.}
    \label{unique-pattern}
\end{figure}

\smallskip
\noindent\textbf{Bugs that only \tool fixes:} Figure~\ref{unique-all} shows the unique bug in Defects4J and the correct fix  that \tool generates, which is equivalent to developers' patch. The correct fix requires ensuring the second parameter to be non-negative.
Pattern-based approaches (e.g., TBar and Hercules) fail to fix it because they have no fix patterns to ensure that a method parameter 
is non-negative. NMT-based approaches (e.g., SequenceR, DLFix, and \coconut) fail to fix it, since such a fix is uncommon. In our patch training data (already 2.72 million training instances from 45,180 projects), there are only two similar fixes. Thus, it is hard for NMT-based models to capture this transformation due to the lack of more similar fixes. However, 
adding ``\texttt{Math.max()}'' to ensure non-negativeness is common in Java methods and is captured by our PL model, allowing \tool to fix the Chart 17 bug in Defects4J correctly.\thibaud{Out of curiosity, I assume this bug is not fixed by BPE+CoNuT+Vanilla, since it require GPT. No need to say it in the paper, but it would be good to check for us.} \nan{right}

As explained in the Introduction, Figure~\ref{fig:uncompilable} shows the KTH bug in QuixBugs, which only \tool fixes and none of the existing techniques evaluated on QuixBugs does. \coconut fails to fix this bug as it generates too many uncompilable patches. Nopol, RSRepair, and Astor cannot repair this bug as they do not implement the required fix pattern.

\todo{update other comparison numbers accordingly 13 more bugs  not 16 unique}
\todo{did DLfix use identifier validity and how? make sure we didnt say we are the first to consider identifier validit}

Comparing with the existing best-performing NMT-based approach \coconut, \tool fixes 
13 more bugs in Defects4J. Figure~\ref{unique-nmt} shows an example bug that \tool fixes and \coconut fails to fix. The correct fix of Closure 10 requires ``\texttt{anyResultsMatch}'', which is nonexistent in the buggy line or context. \coconut prioritizes tokens in the buggy line and context, thus fails to generate the correct token to fix this bug. 
In contrast, \tool's code-aware beam-search strategy extracts all valid identifiers, including ``\texttt{anyResultsMatch}'' which is declared out of the context, and generates the correct fix.

Comparing with the best pattern-based approach, \tool fixes five more bugs in Defects4J than TBar, most of which require complex transformations to fix. Figure~\ref{unique-pattern} shows an example. The correct fix of Math 41 requires changes to the initialization of ``\texttt{i}'' and the condition for the loop. TBar does not have such a complex fix pattern. \tool fixes Math 41 since similar transformations can be learned from the patch training data.

\smallskip
\noindent\textbf{Compilable patch rate:}
In addition to the number of correctly fixed bugs, we use the average compilable rate to measure the effectiveness of \tool learning PL syntaxes and \devlike code. We compare the average compilable rates of the top-k candidate patches generated by different NMT-based models, for bugs in two benchmarks. Table~\ref{tab:compilable rate} shows that \tool generates more compilable patches in top-30 candidates than SequenceR, and more compilable patches in all top-30, 100, 1,000, and 5,000 than \coconut (DLFix does not offer compilable rate data). 
Comparing different rows shows that each component has contributed to the higher compilable patch rate. For example, comparing row ``BPE+GPT+CoNuT+vanilla" with row ``CURE" shows that our code-ware search has increased the average compilable patch rate by 6\% (from 22\% to 28\%) for the top 100 patches.
\tool generates more portions of compilable patches than existing NMT-based approaches, thanks to the PL model and the valid-identifier-check strategy.

\smallskip
These examples and compilable patch rates demonstrate that (1) \emph{the unique capabilities of our model that combines a GPT PL model and an NMT model to learn both developer-like code and fix patterns} to fix more bugs and (2) \emph{the effectiveness of our PL model and the context-aware search strategy in generating more compilable patches.}

\smallskip
\noindent \textbf{Type of bugs that \tool is applicable for:} Similar to most state-of-the-art G\&V APR techniques~\cite{lutellier2020coconut,Li2020dlfix,chen2018sequencer,liu2019tbar,xuan2017nopol,liu2018mining,saha2017elixir,chen2017contract,wen2018context,xin2017leveraging,qi2015analysis,le2012genprog,xiong2017precise,jiang2018shaping,liu2019avatar,koyuncu2018fixminer,le2016history,qi2013does,martinez2018ultra,durieux2016dynamoth,liu2019you}, \tool is designed to fix single-hunk bugs (i.e., the buggy lines and patches are single code segments, and each buggy hunk has separate test cases).

\begin{table}[t]
    \centering
    \caption{Average compilable rates of the top-k candidate patches for bugs in two benchmarks from different models. ``vanilla" and ``code-aware" denote vanilla beam search and code-aware beam search respectively.
    \tool is ``BPE+GPT+CoNuT+code-aware''. 
    `-' indicates data unavailability.} 
    \begin{tabular}{lrrrr}
    \hline
       \textbf{} & \textbf{Top} & \textbf{Top} & \textbf{Top} & \textbf{Top} \\
    \textbf{Model} & \textbf{30} & \textbf{100} & \textbf{1000} & \textbf{5000} \\
    \hline
    \textbf{SequenceR}~\cite{chen2018sequencer} & 33\% & - & - & - \\
    \textbf{\coconut(CoNuT+vanilla)}~\cite{lutellier2020coconut} & 24\% & 15\% & 6\% & 3\% \\
    \textbf{BPE+CoNuT+vanilla} & 28\% & 18\% & 7\% & 4\% \\
    \textbf{GPT+CoNuT+vanilla} & 28\% & 20\% & 9\% & 5\% \\
    \textbf{BPE+GPT+CoNuT+vanilla} & 32\% & 22\% & 10\% & 6\% \\
    \textbf{\tool 
    } & \textbf{39\%} & \textbf{28\%} & \textbf{14\%} & \textbf{9\%} \\
    \hline
    \end{tabular}
    \label{tab:compilable rate}
\end{table}


\subsection{RQ2: What are the contributions of \tool's components?}
\label{rq-contribution}
\noindent
To study the impact of each novel technique  (i.e., GPT PL model, code-aware beam-search strategy, and subword tokenization) of \tool, we compare the following four techniques with \tool: \textbf{CoCoNut (``CoNuT+vanilla'')} An ensemble of ten CoNuT models and ten FConv models, using word-level tokenization and vanilla beam-search strategy.
\coconut uses twice as many models as \tool and the next three techniques (20 versus 10 models) and generates twice as many candidate patches.
Each of the next three techniques is an ensemble of five CoNuT models and five FConv models with the vanilla beam-search strategy. The differences are that \textbf{``BPE+CoNuT+vanilla''} uses subword tokenization, \textbf{``GPT+CoNuT+vanilla''} uses a GPT PL model, and \textbf{``BPE+GPT+CoNuT+vanilla''} uses both subword tokenization and a GPT PL model.

\begin{table}[t]
    \centering
    \caption{Results of ablation study on two benchmarks. (\coconut uses 20 models for ensemble while the rest use only 10 models.)}
    \begin{tabular}{lrr}
        \hline
        \textbf{Model} & \textbf{Defects4J} & \textbf{QuixBugs} \\
        \hline
        \textbf{\coconut(CoNuT+vanilla)} & 44/85 & 13/20 \\
        \textbf{BPE+CoNuT+vanilla} & 45/85 & 16/25 \\
        \textbf{GPT+CoNuT+vanilla} & 44/84 & 19/27 \\
        \textbf{BPE+GPT+CoNuT+vanilla} & 51/94 & 22/27 \\
        \textbf{\tool (BPE+GPT+CoNuT+code-aware)} & \textbf{57/104} & \textbf{26/35} \\
        \hline
    \end{tabular}
    \label{tab:ablation}
\end{table}

All models use a beam size of 1,000, generate 10,000 candidate patches, validate the top 5,000 candidate patches for every bug (except \coconut that generates and validates 20,000 candidate patches for each bug), and are trained on the same dataset. 

\subsubsection{Impact of the GPT PL model}
\label{rq-contribution:gpt}
Table~\ref{tab:ablation} lists the result of the ablation study on two benchmarks. Rows ``BPE+GPT+CoNuT+vanilla'' versus ``BPE+CoNuT+vanilla'' show that the GPT PL model helps APR models fix six more bugs in each benchmark. Comparing ``GPT+CoNuT+vanilla'' with CoCoNuT shows that the GPT PL model helps fix six more QuixBugs bugs. Although they fix the same number of Defects4J bugs, \coconut uses an ensemble of 20 models, while ``GPT+CoNuT+vanilla'' uses only 10. CoCoNuT with 10 models fixes 38 bugs only~\cite{lutellier2020coconut}, which shows an improvement of six more bugs of ``GPT+CoNuT+vanilla'' versus CoCoNuT with 10 models. 


In Table~\ref{tab:compilable rate}, comparing ``BPE+CoNuT+vanilla'' and ``BPE+GPT+CoNuT+vanilla'' shows that GPT increases the average compilable rate by 2\%--4\%. In addition, the average rank (the highest rank one is the best) of the correct patches (before validation) generated by ``BPE+GPT+CoNuT+vanilla'' is 68\% higher than that of ``BPE+CoNuT+vanilla'' (131 vs. 414), indicating that the GPT PL model not only enables APR models to fix more bugs but also improves the ranks of correct patches.

\smallskip
\subsubsection{Impact of Code-Aware Beam-Search Strategy}
\label{rq-contribution:beamsearch}
Comparing ``BPE+GPT+CoNuT+vanilla'' and \tool in Table~\ref{tab:ablation} shows that our code-aware beam-search strategy helps APR models find more correct patches and fix more bugs (six more in Defects4J and four more in QuixBugs).
Comparison between ``BPE+GPT+CoNuT+vanilla'' and \tool in Table~\ref{tab:compilable rate} shows that our code-aware beam search increases the average compilable rate by 3\%--7\%. 
The average rank of the correct patches (before validation) generated by \tool is 21\% higher than ``BPE+GPT+CoNuT+vanilla'' (101 vs. 131), indicating that our new search strategy also increases the rank of correct patches. 

To measure the impact of the length-control strategy, we compare the length of candidate patches generated by ``BPE+GPT+CoNuT+vanilla'' and \tool. For the ``BPE+GPT+CoNuT+vanilla'' model, the average length difference between candidate patches and correct fixes is seven tokens. In contrast, the average length difference between the \tool's candidate patches and correct fixes is five tokens. This shows the length-control strategy helps generate more candidate patches with similar length to the correct fixes. Specifically, it helps fix long bugs (e.g., it fixes the longest bug in QuixBugs that cannot be fixed without length-control strategy) since it searches deeper.

\smallskip
\subsubsection{Impact of subword tokenization}
\label{rq-contribution:subword}
Subword tokenization improves the search space by reducing the size of vocabulary (from 139,423 to 50,057 tokens) and covering more correct fixes. \coconut versus ``BPE+CoNuT+vanilla'' shows that subword tokenization helps fix four more bugs (one in Defects4J and three more in QuixBugs). ``GPT+CoNuT+vanilla'' versus ``BPE+GPT+CoNuT+vanilla'' also shows that subword tokenization helps fix 10 more bugs.

We also compare the number of OOV tokens with different tokenization techniques. With word-level tokenization, 14 bugs contain OOV tokens in our benchmarks (e.g., ``\texttt{binsearch}'' and ``\texttt{charno}''). In contrast, all these OOV tokens are separated into more common tokens when using subword tokenization. This shows that subword tokenization helps reduce the vocabulary size, improve the vocabulary, make the model easier to train, and eventually fix more bugs.

\subsection{Execution Time}

\noindent \textbf{Data extraction:} Downloading and extracting 4.04 million methods from 1,700 projects as our PL training data takes one day. We use \coconut's training data shared on GitHub, which takes five days to extract~\cite{lutellier2020coconut}. Both are a one-time cost. 

\smallskip
\noindent \textbf{Training PL model:}
It takes ten days to pre-train the GPT PL model on four GPUs for five epochs. Since the PL model is trained on large and general data, one programming language only needs one PL model and the PL model can be used to enhance tasks other than APR and does not need retraining. 

\smallskip
\noindent \textbf{Fine-tuning APR models:} It takes 12 days to tune the hyperparameters by training 100 APR models for one epoch. Fine-tuning the top-10 APR models takes, on average, 10.7 hours per model. This is a one time cost as the trained APR models do not need retraining when fixing new bugs.

\smallskip
\noindent \textbf{Cost to fix one bug:} In inference, it takes 2.5 minutes on average for \tool to generate 10,000 candidate patches for one bug using four GPUs. During validation, it takes 16.5 minutes on average to validate one bug. Compared with the state-of-art NMT-based approach~\cite{lutellier2020coconut}, \tool uses fewer models and validates fewer patches, thus \tool is faster and fixes more bugs.

\section{Limitations}

\smallskip
\noindent \textbf{Comparison with previous work:} It is difficult to fairly compare our work with all previous work as they use different training data and FL methods. To be as fair as possible, we use the same training data as \coconut, the state-of-the-art NMT technique, and demonstrate significant improvement on both benchmarks. Some previous work uses different training data, but the selection and extraction of data is also a key component of a technique. In addition, to compare with previous APR techniques, we choose to use perfect localization as it is the preferred comparison method~\cite{liu2019you} and previous work~\cite{liu2020efficiency} evaluated most available APR techniques with perfect FL. 

\smallskip
\noindent \textbf{Generalization to other benchmarks and PL:}
We evaluate \tool on two Java benchmarks, but the approach is neither tied to a specific PL nor a specific benchmark.
\tool is generalizable to other languages by updating the PL parser. The benchmarks we chose are very popular, Defects4J being used to evaluate 25 other APR tools.
In the future, we can also evaluate all APR approaches on recent benchmarks such as Bugs.jar~\cite{saha2018bugs} or Bears~\cite{madeiral2019bears}.

\smallskip
\noindent \textbf{Accuracy of the training sets:}
Since our training and pre-training data extraction is conducted automatically, there is a risk that such data is inaccurate. The training data was extracted in previous work~\cite{lutellier2020coconut} and showed to be reasonably accurate on a random sample. For the pre-training data, we extract all complete functions and some of them might be buggy or incorrect. However, the goal of the pre-training is to learn the syntax of the PL, therefore, we mostly care that the data follows the PL syntax, which is verified since we only keep methods successfully parsed by a Java parser.

\section{Related Work}
\smallskip
\noindent\textbf{Deep Learning for APR:}
Different DL-based APR techniques have been developed to fix bugs~\cite{ding2020patching,dinella2019hoppity,lutellier2020coconut, chen2018sequencer, tufano2018empirical,Li2020dlfix}, compilation issues~\cite{santos2017finding,gupta2017deepfix,mesbah2019deepdelta}, or predict the correctness of generated patches~\cite{tian2020evaluating}.
\tool is different from previous work in three ways. First, our subword-tokenization technique addresses the OOV problem encountered by all NMT-based techniques. Second, \tool integrates a new PL model to the APR models that better learns the syntax of source code. Finally, our new code-aware search strategy chooses only valid identifiers during inference, which helps filter out incorrect patches. As a result, \tool generates more reasonable and compilable patches and outperforms all existing techniques. 

\smallskip
\noindent\textbf{Automatic Program Repair:}
Many APR techniques have been proposed,
which use genetic programming~\cite{le2012genprog}, condition synthesis~\cite{long2015staged,xuan2017nopol,xiong2017precise}, state abstraction~\cite{chen2017contract}, 
heuristics~\cite{xin2017leveraging, asad2019impact,jiang2018shaping,wen2018context},
 human-designed fix patterns~\cite{saha2017elixir,martinez2016astor},  mined fix patterns~\cite{long2017automatic,ocariza2014vejovis,liu2018mining,le2016history,sakkas2020type,liu2018lsrepair}, bytecode mutation~\cite{ghanbari2019practical}, or neural program synthesis~\cite{Kavi2020synthesize}. 
\tool uses a new code-aware NMT approach and fixes more bugs than previous state-of-the-art approaches.

\smallskip
\noindent\textbf{Deep Learning in Software Engineering: }
The software engineering community had applied deep learning to performing various tasks such as source code summarization~\cite{gu2018deep,allamanis2016convolutional},
code clone detection~\cite{white2016deep,li2017cclearner},  defects prediction~\cite{wang2016automatically,li2017software,wang2017software,dinella2019hoppity}, code completion~\cite{fang2020multi}, and
program synthesis~\cite{murali2018neural,ling2016latent}. These techniques, along with ours,  demonstrate 
that deep learning is competitive in different software engineering tasks. Our work introduces code-awareness to DL systems to improve APR. In the future, increasing code-awareness of DL systems applied to other software tasks could also be useful. 

\smallskip
\noindent\textbf{Language Model in Software Engineering: }
Different programming language models have been developed ~\cite{white2015toward,hellendoorn2017deep,allamanis2017survey,chakraborty2018tree2tree,alon2019code2vec,alon2018codeseq,henkel2018code,wang2020reinforcement,hoang2020cc2vec,hu2019code,wang2020blended,wang2020learning,Keim2020}. 
None of these approaches have been evaluated on fixing software bugs and have only been used for simpler tasks such as method name generation or source code summarization. Recent work has questioned the generalizability of some of these approaches for more complex tasks~\cite{kang2019assessing,jiang2019machine}. Compared with these models, \tool uses  GPT~\cite{radford2018improving}, one of the most powerful language models in NLP, to capture code syntax and demonstrates its effectiveness for the more complex APR task.

\section{Conclusion}
\noindent We propose and evaluate \tool, a new NMT-based program repair technique that by design parses, models, and searches source code, as opposed to natural language text, to automatically fix bugs. \tool uses an NMT model that contains a PL model, a code-aware search strategy, and a subword-tokenization technique to create a smaller search space that contains more correct patches and find more correct patches. \tool outperforms all existing techniques on  two popular  benchmarks, fixing 83 bugs. We highlight this direction of code-aware NMT for automatic program repair.

\balance
\bibliographystyle{./bibliography/IEEEtran}
\bibliography{paper}

\end{document}